\documentclass[
 reprint,
 amsmath,amssymb,
 aps,
]{revtex4}

\usepackage{graphicx}
\usepackage{dcolumn}
\usepackage{bm}

\usepackage{fancyhdr} 
\begin{document}


\begin{center}
\textbf{\Large{Theoretical Characterization of the Magnetic Properties of Vanadium-doped Ti\textsubscript{2}C MXenes}}
\vspace{0.2cm}

    Carlos Patiño$^1$, Pablo Díaz$^1$, Nicolás Vidal-Silva$^1$,  Eduardo Cisternas$^1$, Eugenio Vogel$^{1,2}$, Fabian Dietrich$^1$ \\ \vspace{0.1cm}
    $^1${Departamento de Ciencias Físicas, Universidad de La Frontera, Temuco, Chile}\\
$^2${Facultad de Ingeniería y Arquitectura, Universidad Central de Chile, Santiago, Chile}
\end{center}

\hrule
\hspace{0.3cm}

\noindent
\textbf{Abstract} \\
MXenes are two-dimensional materials composed of transition metals and light elements, known for their high conductivity and versatile surface chemistry. The introduction of spin centers via doping can lead to promising materials for spintronics, magnetic sensing, and data storage. We study the effect of V doping in Ti\textsubscript{2}C using first principle and Monte-Carlo simulations. 
 {Our results show that pristine Ti\textsubscript{2}C exhibits ferromagnetic intralayer and antiferromagnetic interlayer exchange interactions, yielding an antiferromagnetic ground state. Vanadium incorporation alters these couplings, yet all doped configurations — Ti\textsubscript{7}VC\textsubscript{4} and three Ti\textsubscript{6}V\textsubscript{2}C\textsubscript{4} variants — retain predominantly antiferromagnetic order. 
Despite the preserved ground state, V doping enhances the magnetic response, most notably in the $p$-Ti\textsubscript{6}V\textsubscript{2}C\textsubscript{4} configuration, which displays increased low-field susceptibility and partial spin alignment across layers. As experimentally isolating individual doped phases is unlikely, samples will consist of mixed configurations whose collective behavior nonetheless exhibits clear signatures of V-induced magnetic modification. These results reveal how transition-metal substitution tunes exchange interactions in MXenes and offer guidance for engineering their magnetic functionalities.}


\hspace{0.1cm}
\hrule

\subsection{Introduction}

The MXene family constitutes a versatile class of two-dimensional materials. Since the first synthesis of the MXene prototype Ti\textsubscript{3}C\textsubscript{2} by exfoliation of the MAX phase Ti\textsubscript{3}AlC\textsubscript{2} in 2011 \cite{Naguib2011}, a lot of research groups investigated properties of these materials varying the composition using (early) transition metals (M), non-metals (X = C, N), and functional groups (T = F, O, OH etc.) for materials having the general formula M\textsubscript{n+1}X\textsubscript{n}T\textsubscript{x} \cite{Anasori2017}. With the large variety of composition, many different applications for these MXenes are reported, \textit{e.g.},  {catalysts for hydrogen reduction reactions \cite{Li2021, Bai2021, ZehtabSalmasi2024}, oxygen evolution reactions \cite{Chen2023, Gandara2024}, CO\textsubscript{2} reduction \cite{Meng2024, Talas2025, Tariq2024}; energy storage devices \cite{Shinde2022, Subramanyam2024, Li2022}, supercapacitors \cite{Chen2022b, Badawi2024, Saju2025}, sensors \cite{Ho2021, Pei2021, Xie2023}, electromagnetic interference shielding \cite{Persson2019, Iqbal2021, Han2020}, electromagnetic absorption \cite{Chen2024, Hao2024, Liu2024}, water purification \cite{Huang2023, Sun2024, Meskher2025}, or biomedical applications \cite{Solangi2023, Chen2022, Lee2024}}.

Among the tunable characteristics of MXenes, their magnetic properties have attracted growing interest in recent years. These properties are strongly influenced by factors such as the choice of transition metal  {(\textit{e.g.} Co \cite{Mokkath2025}, Mn \cite{Zamkova2024}, Cr \cite{Tong2025, Zhang2025}, Fe \cite{Agapov2024})}, the type of surface termination \cite{Thangavelu2025, Sakhraoui2024}, and even the presence of vacancies \cite{Shukla2020}. For instance, theoretical studies have demonstrated that specific surface functionalizations can induce or modify magnetic ordering, making MXenes promising candidates for spintronic applications \cite{Vnosov2024}. However, most of these studies focus on systems incorporating only a single type of transition metal, typically titanium, limiting the accessible magnetic phases and phenomena \cite{Li2022b, Khazaei2012}.  From a magnetic standpoint, Ti\textsubscript{2}C is relatively straightforward to describe. In this titanium carbide, the electronic configuration of the Ti\textsuperscript{2+} ion can be approximated as [Ar]3d\textsuperscript{2}, corresponding to a spin state of $S = 1$. The Ti\textsubscript{2}C structure consists of a carbon layer sandwiched between two layers of titanium atoms, giving rise to the possibility of distinct magnetic coupling patterns. Indeed, previous studies have reported that the magnetic ground state is characterized by ferromagnetic coupling among Ti\textsuperscript{2+} ions within each layer, while the coupling between the two titanium layers is antiferromagnetic \cite{Lv2020, GarcaRomeral2023, GarcaRomeral2023_2}. This arrangement leads to a complete cancellation of the individual magnetic moments, resulting in a net magnetization of zero.
The introduction of small amounts of other transition metal ions into Ti\textsubscript{2}C can create perturbation centers with spin states $S \neq 1$. These altered spin centers may lead to the emergence of magnetic skyrmions, as previously reported \cite{Kulka2023}. One of the simplest and most effective ways to introduce such a perturbation is through vanadium doping  {, which has been experimentally shown for the Ti\textsubscript{3}C\textsubscript{2} MXene \cite{Garg2021, Jin2024, Tian2023}.} Substituting Ti\textsuperscript{2+} with V\textsuperscript{2+} introduces a spin center with $S = \frac{3}{2}$, which does not cancel out under the antiferromagnetic interlayer coupling. As a result, the doped MXene exhibits a net magnetic moment.

Functionalizing 2D materials by doping with transition metals has proven effective, as shown in several studies using vanadium—an element notable for its versatility in reacting with second-row elements and producing vanadium ions with different valences \cite{Demeter2000}. For example, vanadium doping in MoS$_2$ has been shown to significantly improve electrical conductivity while enabling spin-orbit pair functionality, paving the way for efficient magnetization switching in nanoscale devices \cite{Sahoo2025}. Similarly, the Ta$_{0.67}$V$_{0.33}$Se$_2$ alloy exhibits clear signatures of intrinsic two-dimensional ferromagnetism at room temperature, highlighting the stability of magnetic order under ambient conditions \cite{Du2024}. Vanadium-doped WS$_2$ monolayers have also emerged as promising room-temperature dilute magnetic semiconductors, expanding the family of 2D materials suitable for magnetic applications \cite{Zhang2020}. Vanadium incorporation into MoSe$_2$ has been investigated for its ability to introduce magnetic features while maintaining desirable structural characteristics \cite{DelPiero2025}. Furthermore, co-doping approaches, such as the introduction of vanadium and manganese into HfS$_2$ under strain, have revealed complex interactions between structural, magnetic, and optical properties, as demonstrated by first-principles calculations \cite{Lin2024}. These advances demonstrate that vanadium doping could give rise to unexplored magnetic phenomena in MXenes.

Exploring MXenes that incorporate multiple transition metals opens a promising avenue for engineering complex and potentially novel magnetic behaviors. Such systems enable synergistic interactions between different metal species, potentially giving rise to emergent magnetic states not observed in single-metal MXenes. Despite this potential, multi-metallic MXenes remain relatively underexplored—both experimentally and theoretically—representing a valuable opportunity for future research.

In this article, we investigate the effects of vanadium doping in Ti\textsubscript{2}C by determining the exchange coupling constants ($J$) and magnetic anisotropy parameters ($K$) through first-principles calculations. These interaction parameters, along with the corresponding interatomic distances, are used as input for Monte Carlo simulations to study the stability of the magnetic configurations. Finally, we generate hysteresis curves to extract the magnetic properties as a function of the vanadium concentration and temperature.

 \subsection{System}

\begin{figure}[h]
    \centering
    \includegraphics[width=.7\linewidth]{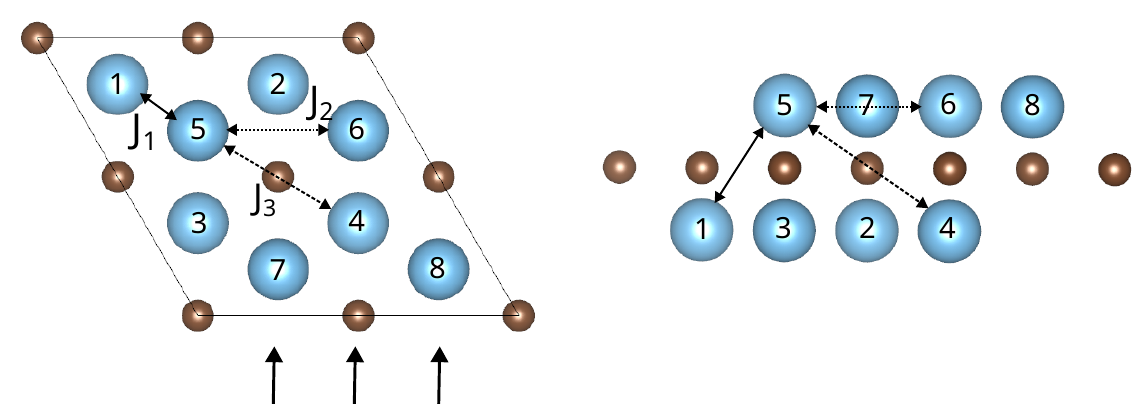}
    \caption{Representation of the 8 magnetic centers in the Ti$_8$C$_4$ supercell and considering the magnetic coupling. Left: top view, right: front view  in the direction of the arrows. Titanium atoms are represented in blue, carbon in brown.}
    \label{Definition}
\end{figure}

Starting with the pristine system Ti\textsubscript{2}C, we constructed a 2x2x1 supercell, resulting in the Ti\textsubscript{8}C\textsubscript{4} cell shown in Figure \ref{Definition}. 

Here, we use the definitions as reported in the literature \cite{Lv2020} to construct the spin Hamiltonian $H_\mathrm{spin}$. The exchange coupling constants $J_i$ can be obtained considering a ferromagnetic (FM) and different anti-ferromagnetic (AFM) cases.

\begin{equation}
    H_\mathrm{spin} = - \sum_{i,j}J_1S_iS_j - \sum_{k,l}J_2S_kS_l - \sum_{m,n}J_3S_mS_n,
\end{equation}
where $J_1$, $J_2$, and $J_3$ represent nearest, next-nearest, and next-next-nearest neighbors, respectively. In the literature \cite{Lv2020, GarcaRomeral2023}, three AFM cases are defined with spin combinations with respect to the atoms shown in Figure \ref{Definition}: A-AFM (1,2,3,4 = $\uparrow$, 5,6,7,8 = $\downarrow$), C-AFM (1,2,5,6 = $\uparrow$, 3,4,7,8 = $\downarrow$) and G-AFM (1,2,7,8 = $\uparrow$, 3,4,5,6 = $\downarrow$).

In the 2x2x1 supercell, the magnetic cases will have the following energies:

\begin{equation}
    E_\mathrm{FM/A-AFM} = E_0 - (\pm12 J_1 + 24 J_2 \pm 12 J_3)|\vec{S}|^2,
\end{equation}

\begin{equation}
    E_\mathrm{C-AFM/G-AFM} = E_0 - (\pm4 J_1 - 8 J_2 \mp 12 J_3)|\vec{S}|^2,
\end{equation}

with $E_0$ representing the electronic energy without considering magnetic interactions. 

Combinations of eqs. (2) and (3) are used to obtain the exchange parameters $J_1$, $J_2$ and $J_3$,  {as shown in detail in the Electronic Supplementary Information (ESI).}

\begin{figure}[h]
    \centering
    \includegraphics[width=.7\linewidth]{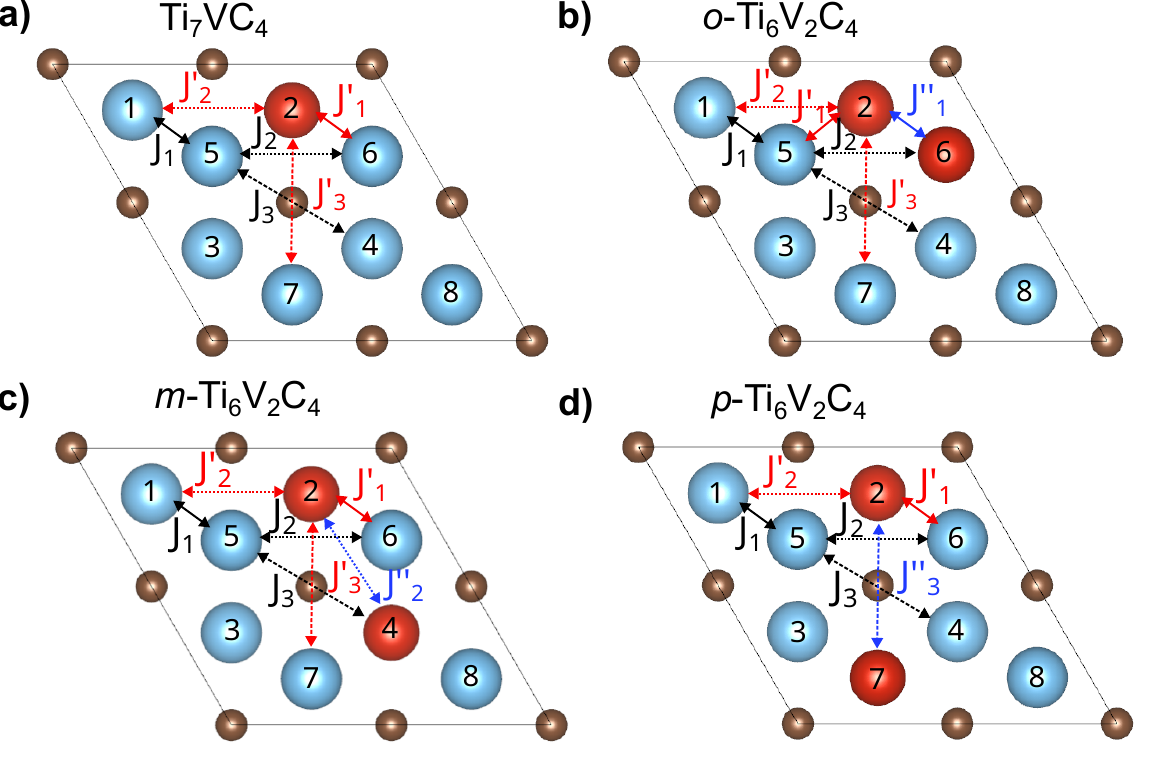}
    \caption{Substitution patterns and new coupling constants in vanadium doped titanium carbide. Vanadium atoms are represented in red. The nomenclature of Ti\textsubscript{6}V\textsubscript{2}C\textsubscript{4} is adapted from organic chemistry, classifying double substitution in a hexagonal arrangement as \textit{ortho} (1,2), \textit{meta} (1,3) or \textit{para} (1,4). }
    \label{TiVC}
\end{figure}

By substituting titanium with vanadium, we expect to create a system that exhibits a net magnetic moment, even in cases of antiferromagnetic coupling. To explore the behavior of two distinct materials and investigate a broader range of coupling constants, we examined the substitution of one or two titanium atoms in the 2x2x1 supercell. This substitution formally results in Ti\textsubscript{7}VC\textsubscript{4} and Ti\textsubscript{6}V\textsubscript{2}C\textsubscript{4}, as illustrated in Fig \ref{TiVC}.

Note that introducing vanadium into the system, demands considering different spin-spin couplings: those between Ti–Ti ($J_i$), Ti–V ($J'_i$), and V–V ($J''_i$). In the case of the Ti\textsubscript{7}VC\textsubscript{4} configuration, only $J_i$ and $J'_i$ are relevant; these couplings are illustrated as black and red arrows, respectively, in Figure \ref{TiVC}a. For the Ti\textsubscript{6}V\textsubscript{2}C\textsubscript{4} configurations, the V–V interaction ($J''_i$) must also be included, and it is indicated as blue arrow in Figure \ref{TiVC}b–d.

To accurately account for the three types of magnetic coupling constants, three distinct substitution patterns are analyzed. Depending on the relative positions of the two V atoms, one of the $J''$ values—$J''_1$, $J''_2$, or $J''_3$—will be relevant, in addition to the Ti–Ti and Ti–V couplings. Details of how these coupling constants ($J'_i$ and $J''_i$) are extracted from various magnetic configurations are provided in the Electronic Supplementary Information (ESI).

 \subsection{Methods}
\subsubsection{First principle calculations}
We applied density functional theory (DFT) using the Vienna Ab initio Simulator Package (VASP) \cite{VASP1996}. The exchange-correlation function was adapted from the generalized gradient approximation (GGA) using the Perdew-Burke-Ernzerhof (PBE) functional \cite{PhysRevLett.77.3865}. For the geometry optimizations, we used an plane-wave cut-off of 520 eV in the PAW potential.  {The energy convergence criteria was set to 10$^{-5}$ eV and the force criteria to 10$^{–4}$ eV/\AA}. All simulations are performed with a Monkhorst-Pack mesh of 24$\times$24$\times$1 in the Brillouin zone. For the correct description of the transition metals, we used a Hubbard (U+J) \cite{Shih2012} correction using for both Ti and V the parameters $U$ = 4.0 eV and $J$ = 0.29 eV. These parameters were adopted from \cite{Lv2020}.  {To eliminate interlayer interactions, a vacuum spacing of 12 \AA{} is introduced perpendicular to the MXene sheets. Furthermore, only minor changes are observed in the electronic structure when spin–orbit coupling (SOC) is included. Therefore, SOC is neglected in the electronic structure calculations. However, it is explicitly taken into account in the evaluation of the magnetic anisotropy energy (MAE). For the MAE calculations, a plane-wave cutoff energy of 600 eV and an energy convergence criterion of 10$^{-6}$ eV are used. For those calculations, we treated exclusively the A-AFM state with starting magnetizations directing in the z-direction.}

\subsubsection{Monte-Carlo simulations}

Once the magnetic parameters are obtained from DFT calculations, we investigated the magnetic behavior induced by vanadium doping through the simulation of magnetic hysteresis curves. The Monte Carlo method, combined with the standard Metropolis algorithm, was employed to solve the stochastic magnetic dynamics using an in-house FORTRAN 90 code. The simulations were carried out on a $30 \times 30 \times 2$ magnetic moment lattice with periodic boundary conditions.

Hysteresis curves were generated over a magnetic field range of $\mu_0 H = -6.0$ to $+6.0$ $\mathrm{T}$ in increments of $0.01$ $\mathrm{T}$, with 10000 Monte Carlo steps performed for each field value. To study thermal effects, calculations were conducted at four different temperatures: $T = 0.1$, 5.0, 10.0, and 15.0 K. For simplification, an average magnetic moment was assigned to the titanium and vanadium atoms, as later summarized in Table \ref{Exchange parameters}.

 \subsection{Results and Disussion}

\subsubsection{First principle simulations}

First, we calculated the relative energies of the nonmagnetic state (calculation of the  non-spin-polarized case), along with the previously mentioned ferromagnetic and three antiferromagnetic configurations. These calculations were performed for both the pristine Ti\textsubscript{2}C and its vanadium-doped derivatives. The energies corresponding to each species and magnetic configuration are summarized in Table \ref{energies},  {the corresponding magnetic moments can be found in the Table S2.}

\begin{table}[ht]
    \centering
    \begin{tabular}{c|c|c|c|c|c}
        & Ti\textsubscript{8}C\textsubscript{4}  & Ti\textsubscript{7}VC\textsubscript{4}  & \textit{m}-Ti\textsubscript{6}V\textsubscript{2}C\textsubscript{4} & \textit{o}-Ti\textsubscript{6}V\textsubscript{2}C\textsubscript{4} & \textit{p}-Ti\textsubscript{6}V\textsubscript{2}C\textsubscript{4}\\ \hline
        NM & 0 & 0 & 0 & 0 & 0 \\
        E
        FM  &  -266  & -379 & -476 & -504 & -458 \\
        A-AFM & -377 & -463 & -538 & -552 & -508\\
        C-AFM & -180 & -279 & -381 & -465 & -399 \\
        G-AFM & -121 & -221 & -435 & -485 & -389     \\   
    \end{tabular}
    \caption{Energies $\Delta E$ of the different magnetic configurations relative to the nonmagnetic state, obtained from DFT calculations. All values are given in meV.}    
    \label{energies}
\end{table}

With respect to the pristine structure, we successfully reproduce the literature values reported by Lv et al. \cite{Lv2020}. For this case, the lowest relative energy is observed for the anti-ferromagnetic state, A-AFM. In this configuration, all spins within a single layer are ferromagnetically aligned, while the two layers exhibit opposite magnetic moments. This alignment results in magnetic compensation and a zero net magnetic moment. About 100 meV higher in energy, the ferromagnetic configuration is found, followed by the other two anti-ferromagnetic configurations (also about 100 meV difference).\\
Introducing vanadium (V) atoms into the Ti\textsubscript{2}C structure alters the stability of the magnetic states. For all magnetic states, the introduction of one V atom into the structure (Ti\textsubscript{7}VC\textsubscript{4}) lowers the energy approximately 100 meV in comparison to the non-magnetic state, while the order of the magnetic configurations is maintained. That means that independent of the vanadium concentration, the anti-ferromagnetic configuration (A-AFM) is always the most stable one.   {Building upon this trend, the introduction of a second V atom to form Ti\textsubscript{6}V\textsubscript{2}C\textsubscript{4} further enhances the stabilization of the magnetic states. In this case, three distinct magnetic configurations must be considered (see Figure~\ref{TiVC} and the last three columns of Tab.~\ref{energies}). The results show that the inclusion of the second V atom provides an additional stabilization of approximately 100 meV relative to the non-magnetic configuration, compared with Ti\textsubscript{7}VC\textsubscript{4}.} 

 {Table~S2 summarizes the magnetic moments obtained from the DFT simulations. For Ti atoms, the values depend strongly on the magnetic configuration. In all ferromagnetic states, the Ti magnetic moments lie between 1.6 and 1.7~$\mu_B$, whereas in the antiferromagnetic states they range from 0.7 to 1.2~$\mu_B$ for pristine Ti\textsubscript{8}C\textsubscript{4}, and between 0.0 and 1.5~$\mu_B$ in the V-doped systems. The variations within the same magnetic configuration in the latter cases arise from symmetry breaking effects. These results suggest that the appropriate spin states for Ti are $S = 0, \frac{1}{2},$ and~1. The discrepancy between the nominal and calculated spin states has been previously discussed in the literature~\cite{GarcaRomeral2023_2}, where it was shown that the inclusion of a Hubbard correction is essential to more accurately describe the correlation effects of the Ti $d$~electrons. Nevertheless, the total magnetizations obtained for the ferromagnetic configurations support the use of $S = 1$ for Ti in the magnetic model. In contrast, the calculated magnetic moments for V atoms range between 2.6 and 2.9~$\mu_B$, corresponding to a spin state of $S = \frac{3}{2}$.}

As indicated above, the calculated energies of each magnetic state can be used to calculate the magnetic exchange-coupling constants. In Table \ref{Exchange parameters}, we summarized the obtained results. The constants $J_i'$ and $J_i''$ are obtained from the single and double vanadium doped MXenes, respectively, assuming that the $J_i$ coupling constants for the Ti-Ti interactions remain unchanged {, see equations in the Supporting Information}. 

\begin{table}[h!]
    \centering
    \begin{tabular}{c|ccc}
       
         $i$ & 1 & 2 & 3 \\ \hline
         $J_i/\mathrm{meV}$ &  -6.5 & 21.3 & -12.0  \\
         $J'_i/\mathrm{meV}$ &  4.7 & 14.2 & -5.0 \\
         $J''_i/\mathrm{meV}$ &  17.4 & 1.9 & 0.6  \\ \hline 
    \end{tabular}
    \caption{ {Simulated magnetic exchange parameters, obtained from the combination of DFT energies of the FM and AFM configurations, assuming $|\vec{S}_\mathrm{Ti}| = 1$, $|\vec{S}_\mathrm{V}| = 3/2$.}}
    \label{Exchange parameters}
\end{table}

 {For the Ti–Ti interactions, $J_2$ (21.3~meV) is the largest, indicating that intralayer coupling dominates and is ferromagnetic, while the interlayer couplings $J_1$ (-6.5~meV) and $J_3$ (-12.0~meV) are weaker and antiferromagnetic. These values are similar to the literature values \cite{Lv2020}.
When substituting one Ti atom with V, the overall magnetic behavior remains largely ferromagnetic: both $J'_1$ and $J'_2$ exhibit positive values (4.7 and 14.2 meV, respectively), indicating ferromagnetic exchange within and between layers, while $J'_3$ changes sign to a weak antiferromagnetic interaction.
In the case of double V doping, the exchange interactions are substantially modified. The strongest coupling corresponds to $J''_1 = 17.4$~meV, evidencing a pronounced ferromagnetic exchange between adjacent layers. In contrast, the intralayer interaction $J''_2 = 1.9$~meV and the next-nearest coupling $J''_3 = 0.6$~meV are considerably weaker, suggesting a reduction in the magnetic correlation within the layers. This evolution of the exchange parameters with increasing V concentration indicates a progressive reinforcement of the interlayer ferromagnetic coupling, accompanied by a suppression of long-range intralayer magnetic interactions.}


In a next step, we used the DFT simulation to obtain information about the magnetic anisotropy. For this, we calculated the total energies along the high symmetry directions: (100), (010), (001), (011), (110), (101) and (111). In most of the investigated MXenes, the (001) axis was found to be the minimum axis, with the magnetic anisotropy energy (MAE) given by:
\begin{equation}
    \mathrm{MAE} = E_{(xyz)} - E_{(001)}
\end{equation}

The MAE are of the six directions are summarized in Tab. \ref{MAE} for the pristine Ti\textsubscript{2}C and its V doped derivatives. These values are calculated for the A-AFM configuration,  {starting with a magnetization exclusively in z direction.}

\begin{table}[h]
    \centering
    \begin{tabular}{c|c|c|c|c|c}
         Spin &  \multicolumn{5}{c}{}\\
        Direction & Ti\textsubscript{8}C\textsubscript{4}  & Ti\textsubscript{7}VC\textsubscript{4}  & \textit{m}-Ti\textsubscript{6}V\textsubscript{2}C\textsubscript{4} & \textit{o}-Ti\textsubscript{6}V\textsubscript{2}C\textsubscript{4} & \textit{p}-Ti\textsubscript{6}V\textsubscript{2}C\textsubscript{4} \\ \hline
        (010) & 149 & 162 & 108 & 209  &  349 \\
        (100) & 149 & 162 & 108 & 209 &  348 \\
        (011) & 74  & 81  & 54 & 106 &  173  \\
        (110) & 149 & 162 & 108 & 210 &  348 \\
        (101) & 75  & 81  & 52 & 104 &  173 \\
        (111) & 99  & 107 & 78 & 145 &  232 \\       
    \end{tabular}
    \caption{ {MAE for different spin directions in V doped MXenes, all values given in $\mu$eV/atom.}}
    \label{MAE}
\end{table}

\begin{figure}[h]
    \centering
    \includegraphics[width=.5\linewidth]{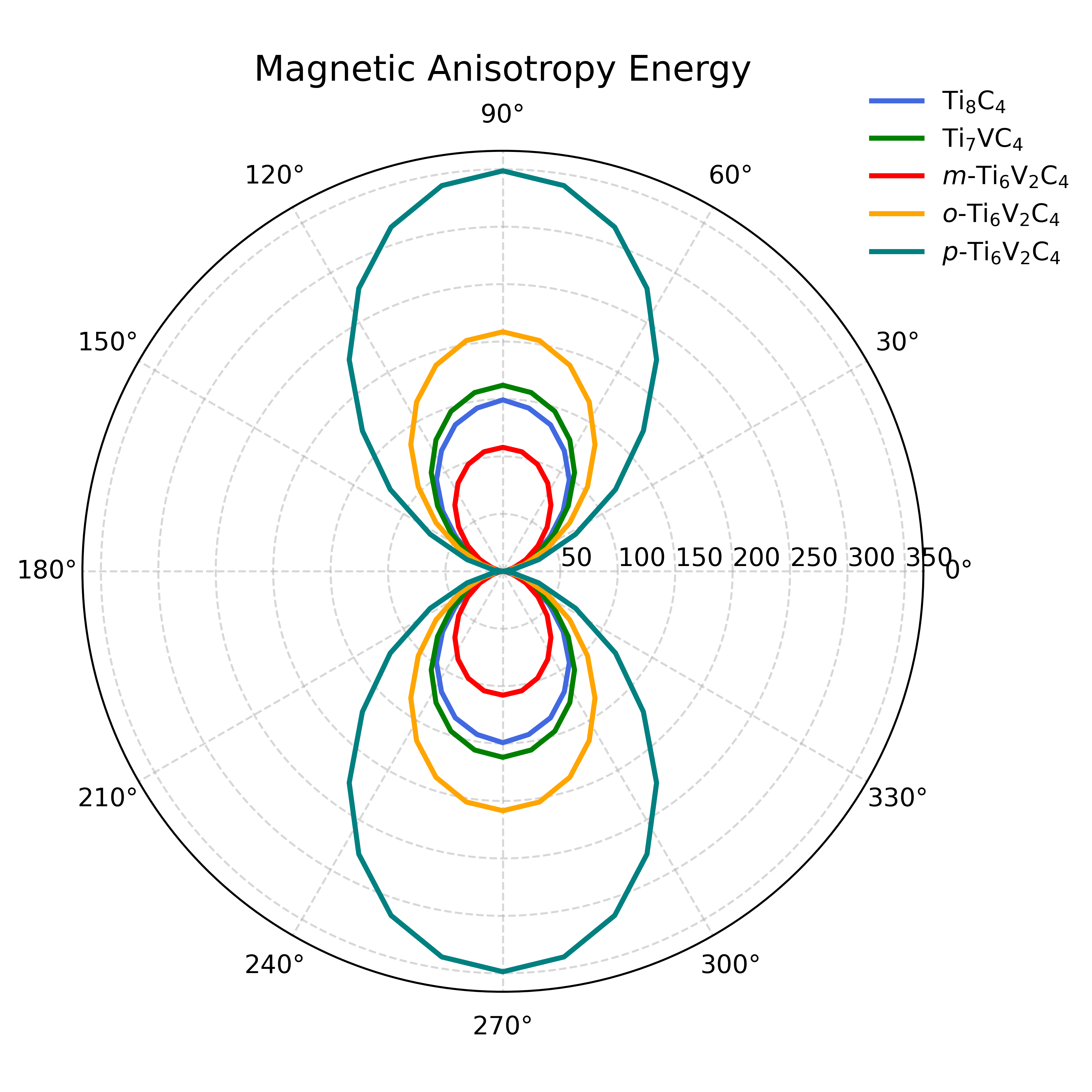}
    \caption{Angle-dependence of the MAE with respect to the rotation of the spin in the yz plane, while 0$^\circ$ represents the (001) direction, values in $\mu$eV/atom.}
    \label{Rotation_PBE+U}
\end{figure}

The DFT calculations show for the Ti\textsubscript{8}C\textsubscript{4}, Ti\textsubscript{7}VC\textsubscript{4} and \textit{p}-Ti\textsubscript{6}V\textsubscript{2}C\textsubscript{4} a regular distribution. While the (001) axis is the minimum direction, the (010), (100) and (110) axes have the same energy, whose value is double the value of the (011) and (101) axes. That means, these materials are isotropic in the xy plane. The absolute values for Ti\textsubscript{8}C\textsubscript{4} and Ti\textsubscript{7}VC\textsubscript{4} are almost identical, but significantly smaller values are obtained for \textit{p}-Ti\textsubscript{6}V\textsubscript{2}C\textsubscript{4}. The other structures \textit{m}-Ti\textsubscript{6}V\textsubscript{2}C\textsubscript{4} and \textit{o}-Ti\textsubscript{6}V\textsubscript{2}C\textsubscript{4} show a certain anisotropy also in the xy plane. While for the \textit{o}-Ti\textsubscript{6}V\textsubscript{2}C\textsubscript{4}, the deviation is quite small, we observe a strange behavior for \textit{m}-Ti\textsubscript{6}V\textsubscript{2}C\textsubscript{4}: the anisotropy is rotated, having the (010) axis as minimum direction. This behavior might be associated with the anti-ferromagnetic coupling of the two V atoms in the same layer, changing the anisotropy of the material.\\
As in the majority of the investigated MXenes, the simple axis is (001) and the xy plane is isotropic, we investigated the angle-dependence of the MAE by rotating the spin in the yz plane by steps of 9$^\circ$, defining the corresponding vectors in the spin-polarized collinear calculations,  {see Table S1 in the Supplement Information.}.
The results are shown in Figure \ref{Rotation_PBE+U}.

These angle-dependent values of the MAE are fitted to obtain the anisotropy constants $K_i$ \cite{Wang2019}:
\begin{equation}
    \mathrm{MAE}(\theta) = K_1\sin^2\theta + K_2\sin^4\theta
\end{equation}

With that, we obtained the following values of the anisotropy parameters $K_1$ and $K_2$ for the investigated species, as summarized in Table \ref{Anisotropy}.

\begin{table}[h]
    \centering
     \begin{tabular}{c|c|c|c|c|c}
         \multicolumn{1}{c}{}& \\
         &Ti\textsubscript{8}C\textsubscript{4}  & Ti\textsubscript{7}VC\textsubscript{4}  & \textit{m}-Ti\textsubscript{6}V\textsubscript{2}C\textsubscript{4} & \textit{o}-Ti\textsubscript{6}V\textsubscript{2}C\textsubscript{4} & \textit{p}-Ti\textsubscript{6}V\textsubscript{2}C\textsubscript{4} \\ \hline
         $K_1$/$\mu$eV/atom & 148.5 & 161.2 & 110.2 & 208.4 & 346.4 \\
         $K_2$/$\mu$eV/atom & -0.12 & 0.54  & -2.31 & 0.06 & 2.1 \\
    \end{tabular}
    \caption{Anisotropy constants obtained by fitting with equation (5).}
    \label{Anisotropy}
\end{table}

\subsubsection{Magnetic properties}

\begin{center}
    \underline{Ground state}
\end{center}

\begin{figure}[h]
    \centering
    \includegraphics[width=.6\linewidth]{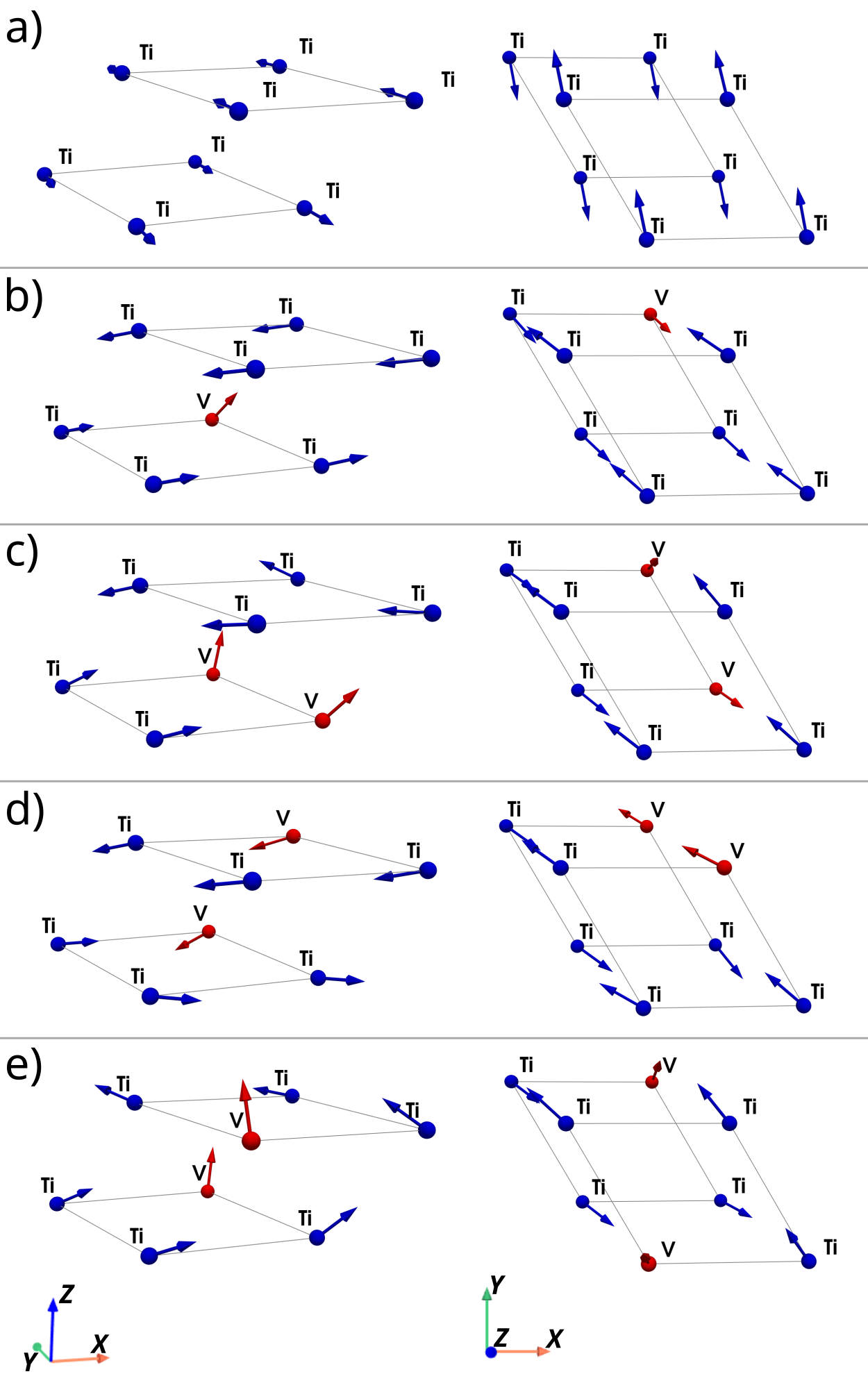}
    \caption{Magnetic configuration in the ground state of the unitary cell for the five configurations studied: a) Ti\textsubscript{8}C\textsubscript{4}, b) Ti\textsubscript{7}VC\textsubscript{4}, c) \textit{m}-Ti\textsubscript{6
    }V\textsubscript{2}C\textsubscript{4}, d) \textit{o}-Ti\textsubscript{6}V\textsubscript{2}C\textsubscript{4}, and e) \textit{p}-Ti\textsubscript{6}V\textsubscript{2}C\textsubscript{4}. The figures on the left correspond to a semi-lateral perspective, and the figures on the right to a top perspective. For each case, the total moment of the supercell normalized to the saturated case is reported. Note: The figure shows that the spin orientations do not coincide with the out-of-plane direction indicated by the effective anisotropy; this is because the anisotropy is at least two orders of magnitude smaller than the exchange interactions, which, from a dynamic point of view, only leads to the minimization of the exchange energy.}
    \label{coercivity_retentivity}
\end{figure}

Equipped with the exchange couplings and the corresponding magnetic anisotropy constants, we are now in a position to explore a system consisting of a larger number of spins. By performing Monte Carlo simulations, we are able to compute the (classical) magnetic ordering of the explored V-doped magnetic MXenes as well as their static properties. Figure \ref{coercivity_retentivity} shows the ground state of Ti\textsubscript{8}C\textsubscript{4} and the four vanadium-doped configurations.

{ In the pristine case (Figure \ref{coercivity_retentivity}a), the intralayer interactions are ferromagnetic ($J_2 = 21.3$ meV), while the interlayer interactions are antiferromagnetic ($J_1 = -6.5$ meV and $J_3 = -12.0$ meV). As a result, the spin configuration shown in Figure \ref{coercivity_retentivity}a emerges: spins within each layer are aligned, whereas spins between layers are anti-aligned. Additionally, the magnetic anisotropy causes the spins to orient out of plane, as expected from the anisotropy constants calculated for this system.

In the Ti\textsubscript{7}VC\textsubscript{4} case (Figure \ref{coercivity_retentivity}b), the Ti atom at position 2 is replaced by a V atom. The Ti–V interaction across layers becomes $J'_1 = 4.7$ meV, transitioning from antiferromagnetic in the case of Ti-Ti ($J_1 = -6.5$ meV) to ferromagnetic, weakening the antiferromagnetic interaction between layers. Within the same layer, the Ti–V coupling decreases to $J'_2 = 14.2$ meV, weakening the intralayer ferromagnetism. The diagonal interlayer interaction decreases from $J_3 = -12.0$ meV (Ti–Ti) to $J'_3 = -5.0$ meV (Ti–V), becoming weakly antiferromagnetic. As depicted in Figure \ref{coercivity_retentivity}b, although these changes are important, they are not strong enough to significantly alter the ground-state spin configuration compared to the pristine case; only a change in the spin orientation of the vanadium atom was observed.

In the \textit{m}-Ti\textsubscript{6}V\textsubscript{2}C\textsubscript{4} system (Figure \ref{coercivity_retentivity}c), Ti atoms at positions 2 and 4 (same layer) are substituted by V atoms. The interlayer Ti–V interaction becomes $J'_1 = 4.7$ meV, weakening the antiferromagnetic coupling between layers. The intralayer Ti–V interaction decreases to $J'_2 = 14.2$ meV, weakening ferromagnetic coupling. The direct interaction between the two V atoms becomes weakly ferromagnetic ($J''_2 = 1.9$ meV), in contrast to the original strong ferromagnetic Ti–Ti coupling. The diagonal interaction between layers changes from $J_3 = -12.0$ meV to $J'_3 = -5.0$ meV, while remaining antiferromagnetic. As shown in Figure \ref{coercivity_retentivity}c), this results in the spins of both vanadium atoms being outside the plane.

For \textit{o}-Ti\textsubscript{6}V\textsubscript{2}C\textsubscript{4} (Figure \ref{coercivity_retentivity}d), the V atoms occupy positions 2 and 6 in different layers. Again, the Ti–V interlayer interaction is $J'_1 = 4.7$ meV, contrary to the antiferromagnetic interaction for Ti-Ti, weakening the antiferromagnetism between the layers. Intralayer Ti–V interactions decrease to $J'_2 = 14.2$ meV, weakening ferromagnetic alignment within each layer. The diagonal Ti–V interaction changes from $J_3 = -12.0$ meV to $J'_3 = -5.0$ meV. A key distinction in this configuration is the direct V–V coupling across the layers, which is ferromagnetic ($J''_1 = 17.4$ meV), in contrast to the antiferromagnetic Ti–Ti coupling, with a magnitude approximately three times larger. This change weakens the antiferromagnetic order between layers. As illustrated in Figure \ref{coercivity_retentivity}d), the alignment between the vanadium atoms is clearly visible.

In the \textit{p}-Ti\textsubscript{6}V\textsubscript{2}C\textsubscript{4} case (Figure \ref{coercivity_retentivity}e), V atoms replace Ti at positions 2 and 7 (in different layers). The interlayer Ti–V interaction becomes ferromagnetic ($J'_1 = 4.7$ meV), and the intralayer Ti–V interaction decreases to $J'_2 = 14.0$ meV, consistent with previous cases. The key difference lies in the V–V interaction along the diagonal, which changes from antiferromagnetic ($J_3 = -12.0$ meV for Ti–Ti) to weak ferromagnetic ($J''_3 = 0.6$ meV for V–V). This important change leads to a partial alignment of the V spins across layers, as seen in Figure \ref{coercivity_retentivity}e, and causes a slight reorientation of neighboring Ti spins compared to the undoped configuration.}

\begin{center}
    \underline{Hysteresis curves}
\end{center}

\begin{figure}[h!]
    \centering
    \includegraphics[width=0.45\linewidth]{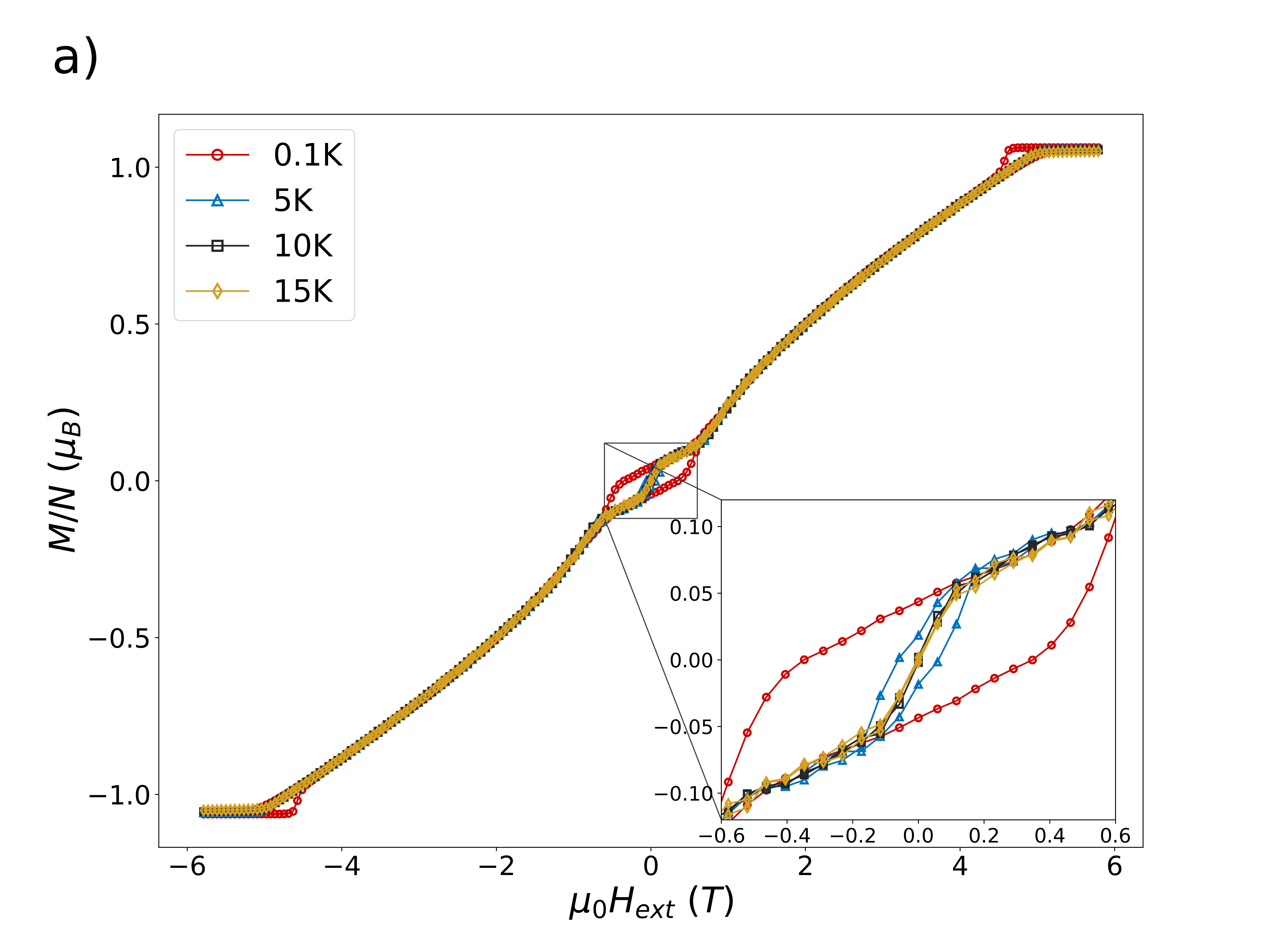}
    \includegraphics[width=0.45\linewidth]{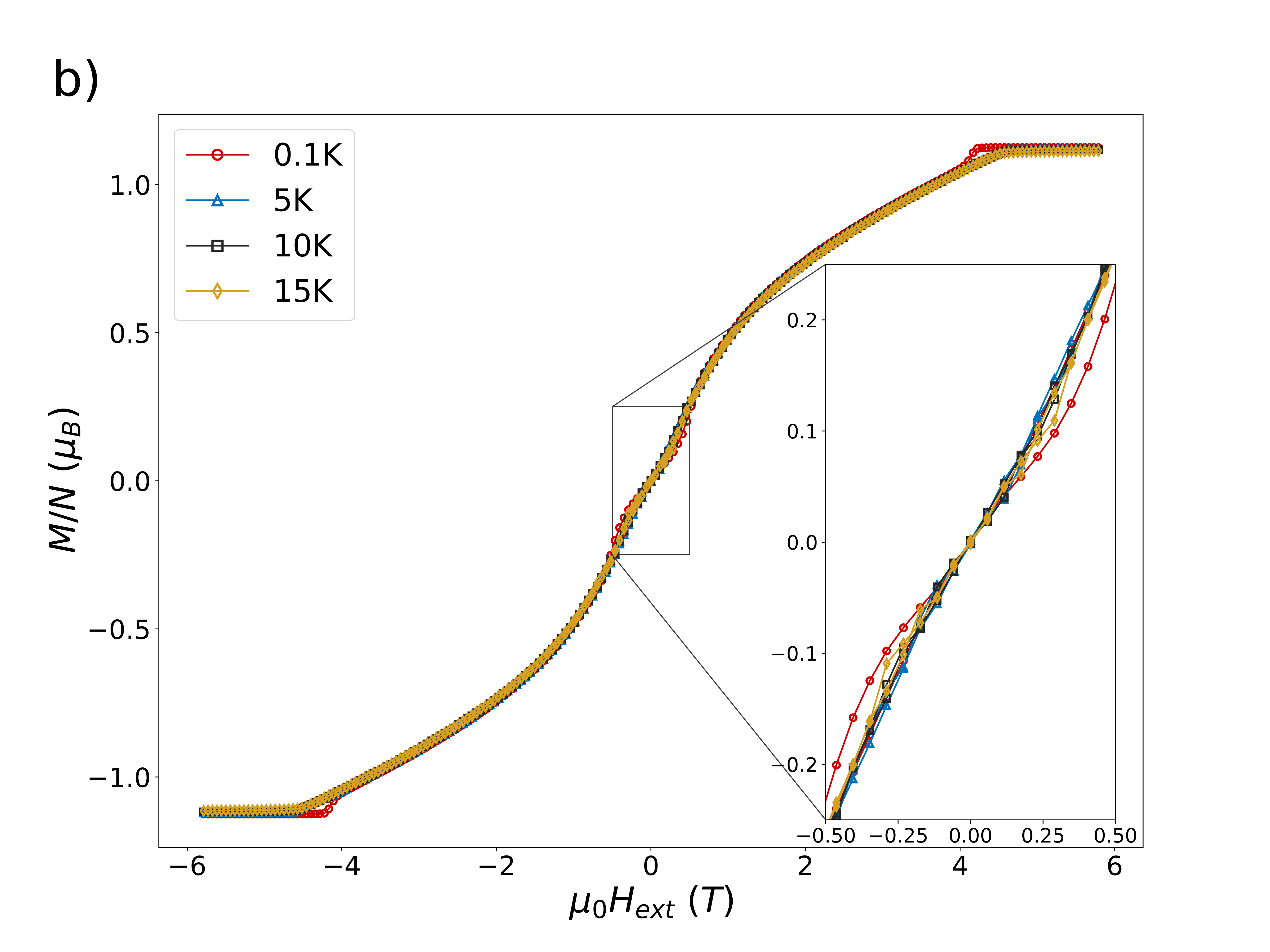}
    \includegraphics[width=0.45\linewidth]{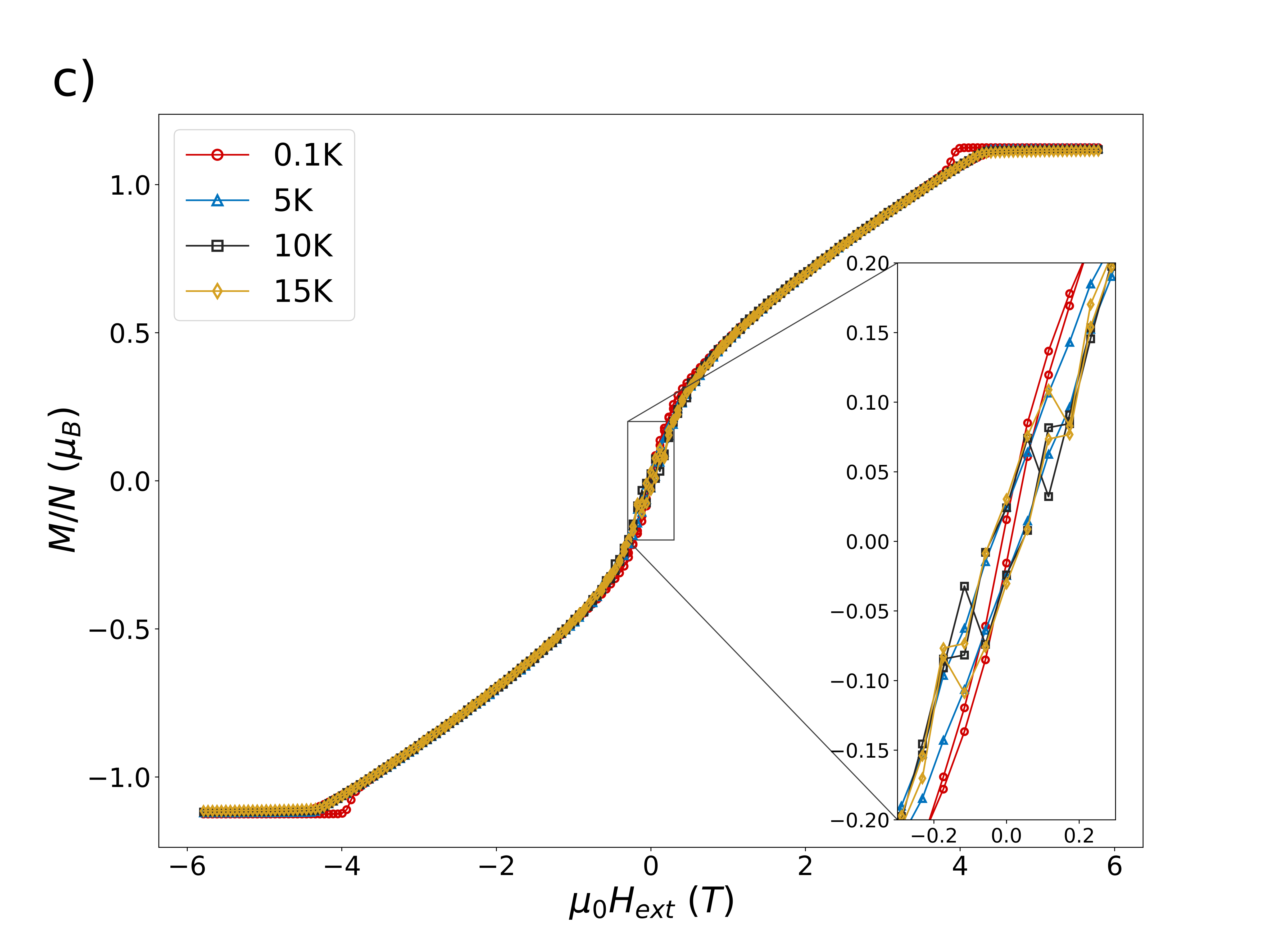}
    \includegraphics[width=0.45\linewidth]{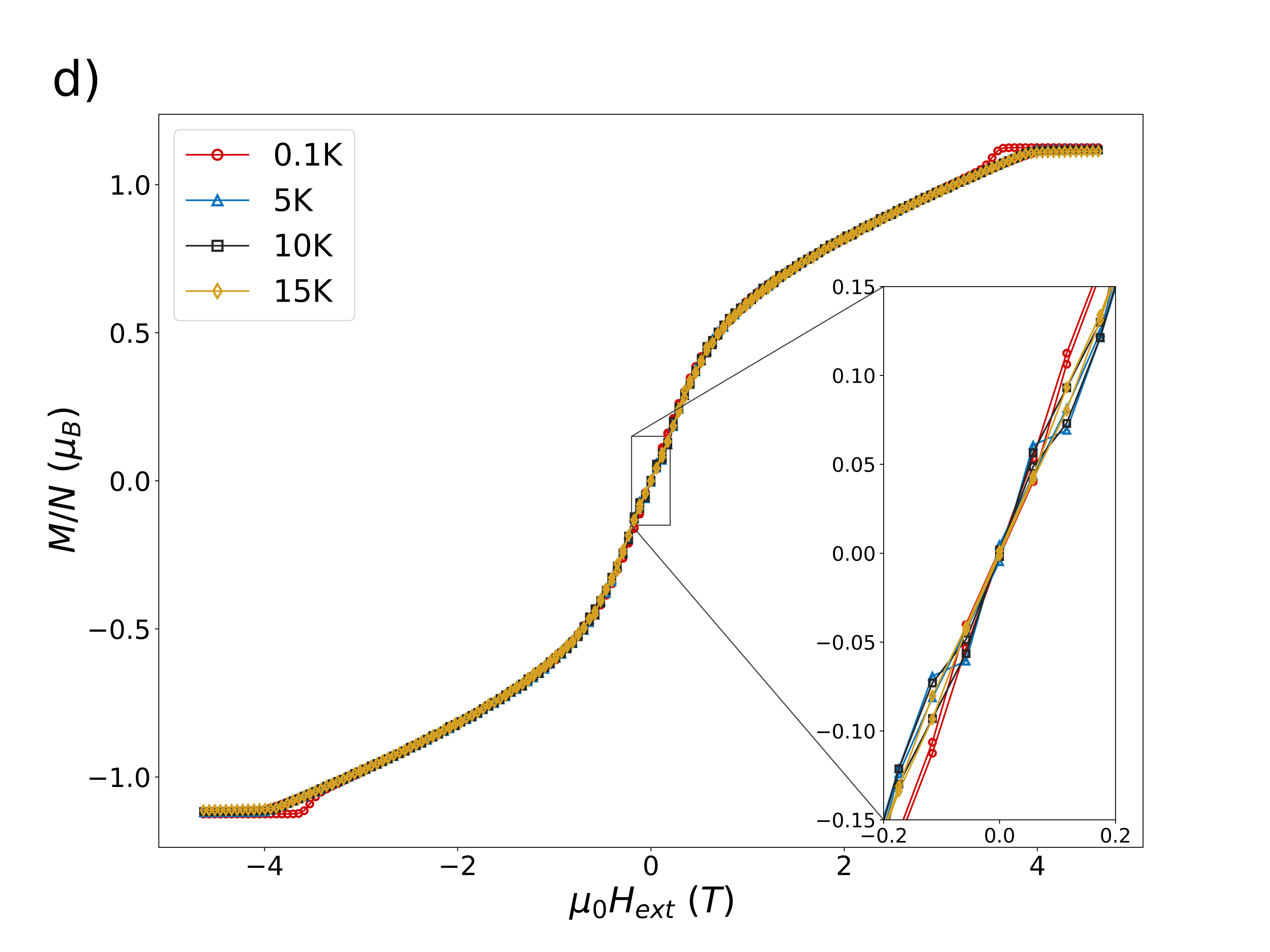}
    \caption{Hysteresis curves for the four vanadium-doped cases: a) Ti\textsubscript{7}VC\textsubscript{4}, b) \textit{m}-Ti\textsubscript{6}V\textsubscript{2}C\textsubscript{4}, c) \textit{o}-Ti\textsubscript{6}V\textsubscript{2}C\textsubscript{4}, and d) \textit{p}-Ti\textsubscript{6}V\textsubscript{2}C\textsubscript{4}. Simulations were performed considering $30\times30\times2$ metal atoms (Ti and {V}) with periodic boundary conditions.}
    \label{hysteresis}
\end{figure}

{  Once the magnetic ground state is calculated, we now focus on the static magnetic properties. Hysteresis curves were obtained for the four vanadium-doped configurations. As a reference, we note that the pristine system, Ti\textsubscript{8}C\textsubscript{4}, exhibits antiferromagnetic behavior, as discussed previously: interlayer interactions ($J_1 = -6.5$ meV and $J_3 = -12.0$ meV) are antiferromagnetic, while intralayer interactions are ferromagnetic ($J_2 = 21.3$ meV). The corresponding hysteresis loop is shown in Figure 2 of the Supplementary Information. The results for the four doped cases are presented in Figure \ref{hysteresis}.

Figure \ref{hysteresis}a shows the hysteresis curve for Ti\textsubscript{7}VC\textsubscript{4}, where one V atom replaces one single Ti atom. Ti-Ti (Ti-V): interlayer interactions $J_1 = -6.5$ meV ($J'_1 = 4.7$ meV) and $J_3 = -12.0$ meV ($J'_3 = -5.0$ meV); intralayer interaction $J_2 = 21.3$ meV ($J'_2 = 14.2$ meV). At very low temperature ($T = 0.1$ K), the system displays ferromagnetic behavior, with a coercive field exceeding $0.5$ T. The calculated value for the magnetizations is around 0.05 $\mu_\mathrm{B}$ per metal atom, which is equivalent to 5.18 Am\textsuperscript{2}/kg, a value at least two  magnitudes larger than experimentally reported values of surface functionalized Ti\textsubscript{3}C\textsubscript{2}T\textsubscript{x} \cite{Scheibe2019} or Nb-doped Ti\textsubscript{3}C\textsubscript{2} \cite{Fatheema2020}. However, as the temperature increases, this behavior rapidly degrades: by $T = 5.0$ K, the hysteresis loop already shows a pronounced collapse, indicating a transition to antiferromagnetic behavior, which persists at higher temperatures.  In summary, a single vanadium does not appreciably change the antiferromagnetic behavior of the undoped system at high temperatures.

Figures \ref{hysteresis}b, \ref{hysteresis}c, and \ref{hysteresis}d show that for \textit{m}-Ti\textsubscript{6}V\textsubscript{2}C\textsubscript{4}, \textit{o}-Ti\textsubscript{6}V\textsubscript{2}C\textsubscript{4}, and \textit{p}-Ti\textsubscript{6}V\textsubscript{2}C\textsubscript{4}, respectively, the antiferromagnetic behavior is quite clear. Only at very low temperatures are some disturbances observed.}

\begin{center}
  \underline{Magnetic susceptibility}  
\end{center}

\begin{figure}[h!]
    \centering
    \includegraphics[width=0.45\linewidth]{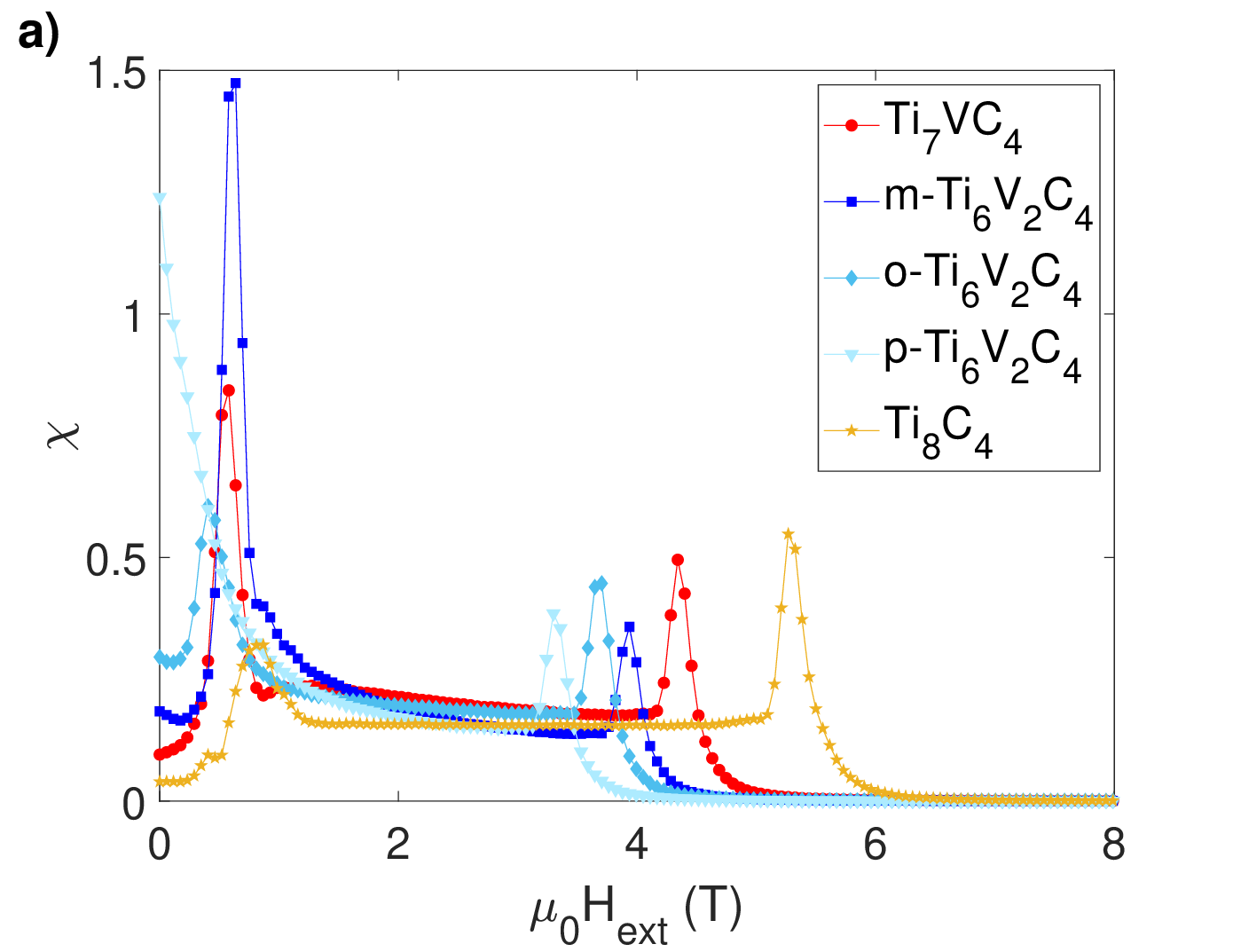}
    \includegraphics[width=0.45\linewidth]{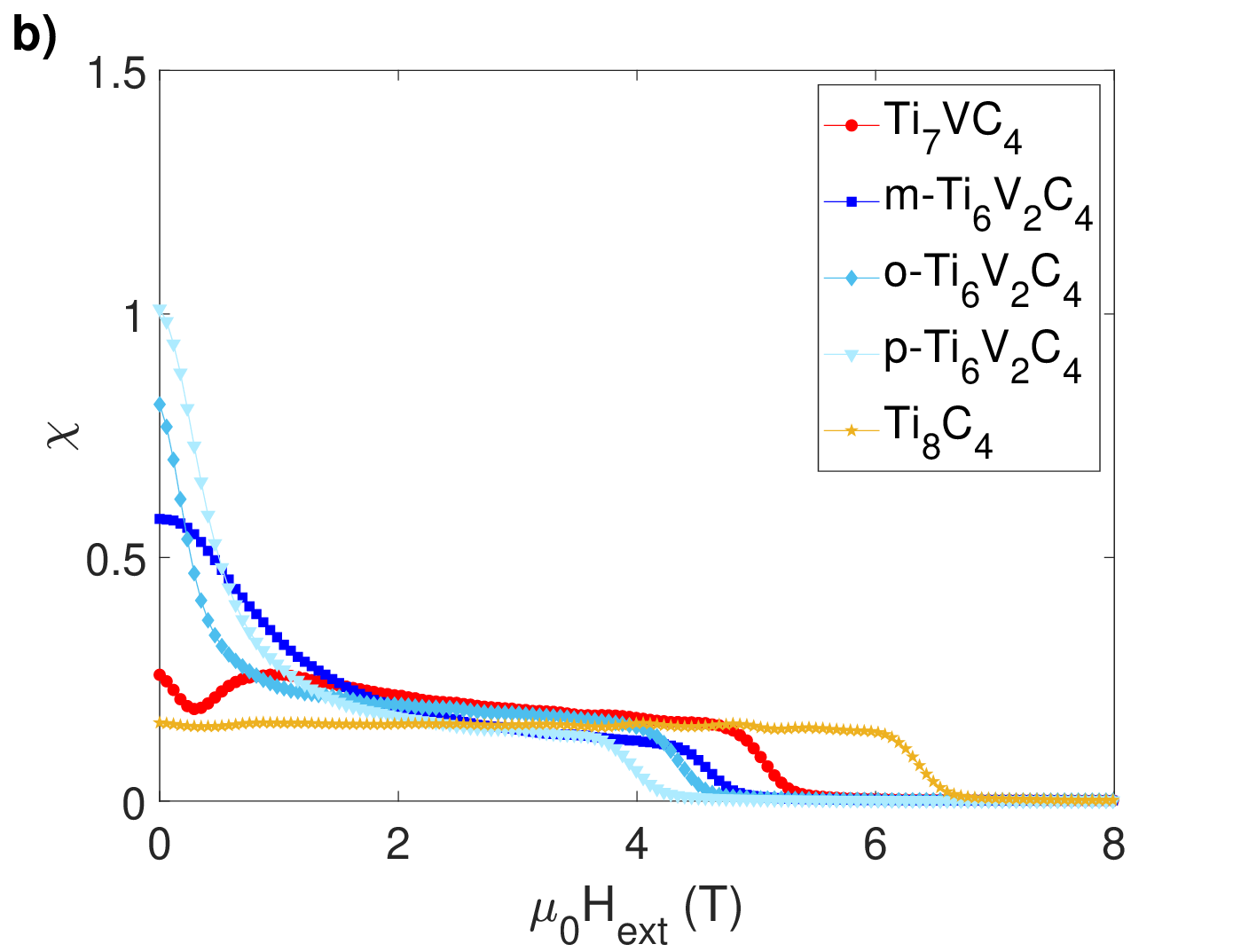}
    \includegraphics[width=0.45\linewidth]{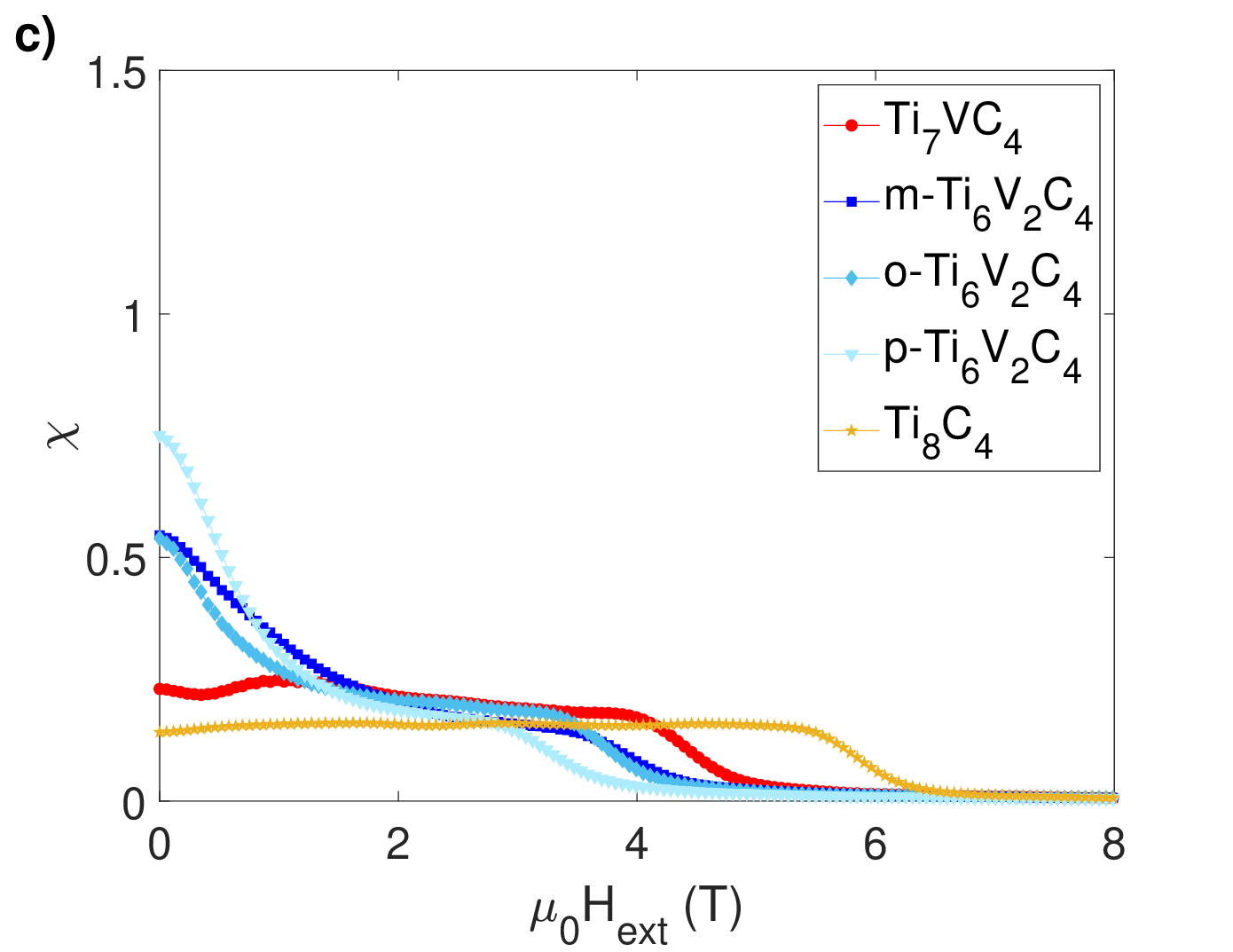}
    \includegraphics[width=0.45\linewidth]{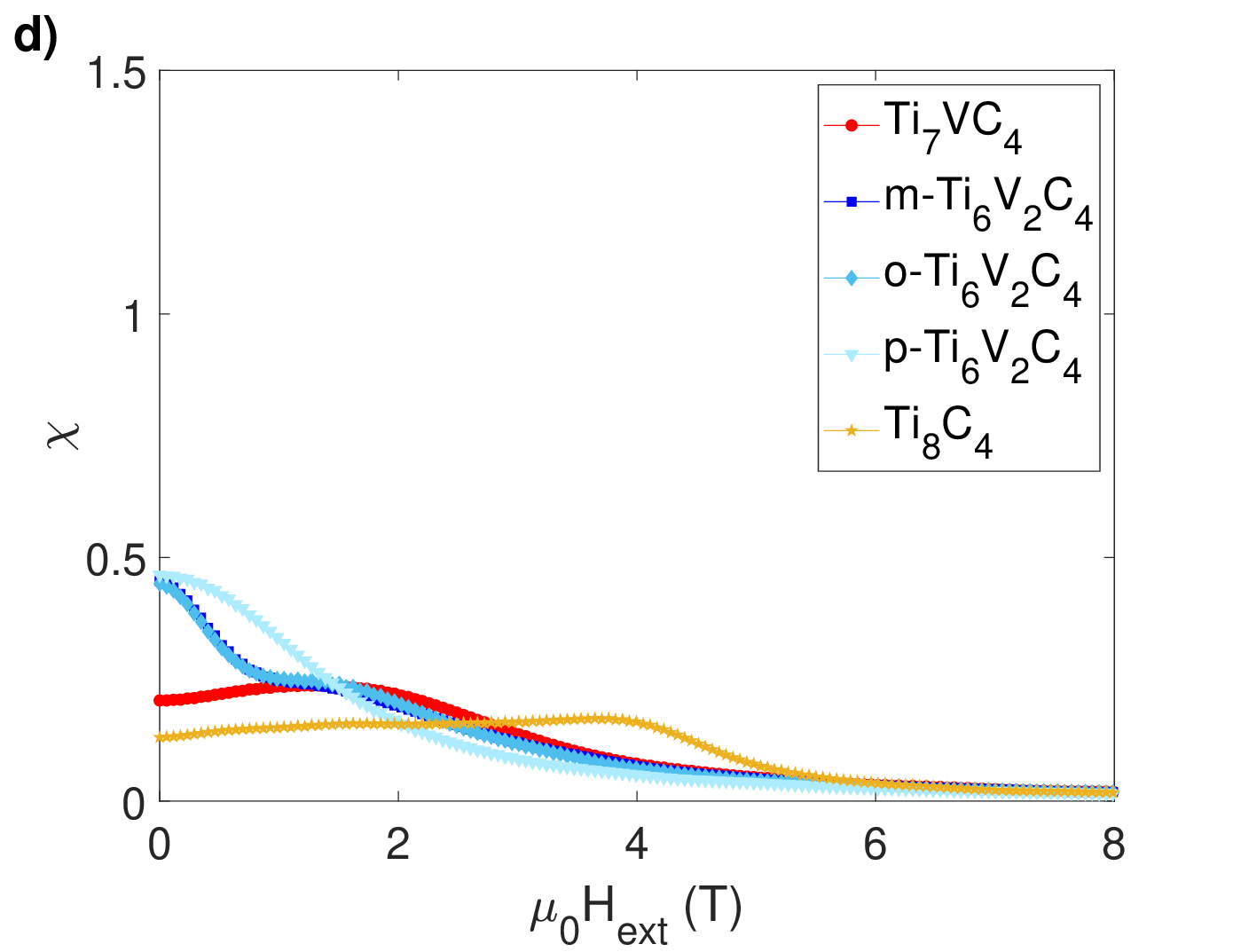}
    \caption{Magnetic susceptibility of doped systems: a) $T=0.1$ K, b) $T=10.0$ K, c) $T=100.0$ K, and d) $T= 300$ K.}
    \label{susceptibility}
\end{figure}

{  As previously noted, doped Ti\textsubscript{8}C\textsubscript{4} exhibits predominantly antiferromagnetic behavior, although the incorporation of vanadium significantly modifies the underlying magnetic interactions. To further elucidate the differences among the systems, we computed the magnetic susceptibility at temperatures ranging from cryogenic values up to near room temperature.

The magnetic susceptibility, $\chi$, was obtained from magnetization curves $M(H)$ calculated by performing independent simulations for each value of the applied magnetic field $H$, starting from random initial magnetic configurations to ensure unbiased convergence toward equilibrium. The resulting $M(H)$ curves are presented in Figure~S3 of the Supplementary Information. The magnetic susceptibility was calculated as $\chi = \mathrm{d}M/\mathrm{d}H$ for each configuration and temperature. The resulting susceptibility curves are illustrated in Figure~\ref{susceptibility}.

Figure~\ref{susceptibility}(a) displays $\chi$ at $T = 0.1$ K. For the pristine system and all doped configurations, two pronounced peaks are observed, which are associated with spin-flop and saturated regime transitions, respectively,—a hallmark of certain antiferromagnetic systems \cite{Bogdanov2007}. Hereafter, we refer to the low-field feature (below 1 T) as peak~1, and to the higher-field feature (between 3 and 6 T) as peak~2. Notably, vanadium doping enhances the amplitude of peak~1 compared to the pristine case, with the most pronounced response found for the \textit{m}-Ti\textsubscript{6}V\textsubscript{2}C\textsubscript{4} configuration. This enhancement reflects a stronger magnetic response induced by doping, arising from the modification of Ti–Ti antiferromagnetic interactions into partially ferromagnetic Ti–V or V–V exchange pathways.

Figures~\ref{susceptibility}(b), \ref{susceptibility}(c), and \ref{susceptibility}(d) show the susceptibility for $T = 10$, 100, and 300 K, respectively. In all three cases, the spin-flop– and saturated-related peaks disappear as thermal fluctuations overcome the field-induced reorientation processes. For the pristine system, $\chi$ remains approximately constant up to magnetic saturation, which occurs at fields close to 6 T. In contrast, the Ti\textsubscript{7}VC\textsubscript{4} system exhibits a noticeable reduction in $\chi$ in the low-field region where peak~1 was previously observed at low temperature. Although all doped systems present a higher susceptibility than the pristine case, $\chi$ drops to zero at lower applied fields due to the earlier onset of magnetic saturation.

The doubly doped configurations exhibit a clearly enhanced susceptibility for fields below 2 T, reaching a maximum for the \textit{p}-Ti\textsubscript{6}V\textsubscript{2}C\textsubscript{4} system. These results demonstrate that vanadium doping significantly increases the sensitivity of the system to an external magnetic field, an effect that persists up to room temperature.

Our results indicate that doping with two vanadium atoms per supercell produces the most distinct magnetic behavior compared to the pristine case. However, from an experimental standpoint, it is not possible to guarantee that doping alone will always result in two vanadium atoms per supercell. While experimentally isolating each of these doped phases may present considerable challenges, a material composed of a mixture of all possible configurations is expected to exhibit a noticeable change in magnetic behavior, as discussed in relation to magnetic susceptibility. This is better appreciated if the abundance of V presence in the supercell configurations is followed as the V concentration $x$ increases. In Figure S4, we present the result of supercells with an increasing number of random occupation by V atoms up to $x=0.1$. It is clear that the case of the magnetic cell Ti\textsubscript{7}VC\textsubscript{4}, with one V atom, dominates the doped cells. It is followed by the cells with two V atoms, of which two are magnetic. Supercells with higher V occupancy are not important for this range of doping concentrations. Higher concentrations would require a different approach oriented to alloys rather than doping, as is the case in the present study. }


 \subsection{Conclusions}

In this work, we have investigated the magnetic properties of vanadium-doped Ti\textsubscript{2}C MXenes by combining first-principles calculations with Monte Carlo magnetic simulations. The calculated exchange coupling constants and magnetic anisotropy parameters reveal that even low concentrations of vanadium significantly modify the magnetic landscape, breaking the perfect interlayer compensation found in pristine Ti\textsubscript{2}C and inducing a net magnetization. Our Monte Carlo results show that the incorporation of vanadium does not change the antiferromagnetic behavior of the system, but it significantly changes the magnetic susceptibility, mainly to fields less than 2 T, with the change being greater when the doping is double per unit supercell. These findings indicate that interlayer coupling can be effectively tuned via targeted transition-metal substitution.

Beyond providing a deeper understanding of the microscopic interactions in doped MXenes, our results point toward the feasibility of engineering two-dimensional materials with tailored magnetic responses for spintronic. The demonstrated sensitivity of the magnetic ground state to vanadium concentration suggests a rich design space for achieving nontrivial spin textures, such as skyrmions, in related MXene systems. Future studies may explore the combined effects of multi-metal doping, strain engineering, and different surface terminations to further expand the functional versatility of these promising materials.

 \subsection*{Conflict of interest}
All authors declare that they have no conflicts of interest.

 \subsection*{Data availability statement}
The data of this research, including input and output data of DFT and Monte Carlo simulations, are available at ZENODO repository at https://doi.org/10.5281/zenodo.17571790.

 \subsection*{Ethical statement}
The present study is based solely on theoretical simulations  and does not involve human participants, animal subjects, or any data derived from them. Accordingly, ethical approval from an institutional or national research ethics committee was not required.

 \subsection*{Acknowledgement}
 {Powered@NLHPC: This research was partially supported by the supercomputing infrastructure of the NLHPC (CCSS210001)}

 \subsection*{Funding}
The authors thank the Agencia Nacional de Investigación y Desarollo (ANID) for financial support of this research in project FONDECYT de Iniciación 11230223 (F.D.), FONDECYT Regular 1231020 (P.D.), 1230055 (E.V.), and 1250364 (N. V-S.). E.V. also thanks Centro de Investigación Asociativa ANID CIA25002. E.C thanks the Dirección de Investigación of the Universidad de La Frontera DIUFRO (DI23-0026) for financial support. E.C., P.D. and F.D. acknowledge partial financial support from the project “Implementación de una unidad interdisciplinar para el desarrollo de Tecnologías Aplicadas y Ciencias (InTec)”, Code “FRO2395”, from the Ministry of Education of Chile.

 \subsection*{Author contribution}
Conceptualization: F.D., P.D., N.V.; Methodology: C.P, F.D., P.D.; Software: C.P., F.D., P.D.; Validation: N.V., P.D., F.D.; Formal analysis: C.P., P.D., F.D.; Investigation: C.P., P.D., F.D.; Data curation: F.D., P.D., Writing -- original draft: F.D., N.V., P.D., E.C., E.V.; Writing -- Review $\&$ Editing: F.D., N.V., P.D., E.C., E.V.; Visualization: C.P., F.D., Supervising: F.D., P.D., N.V.; Project administration: F.D., P.D., N.V.; Funding acquisition: F.D., P.D., N.V., E.C., E.V. 

\bibliographystyle{ieeetr}
\bibliography{MXenes}

\end{document}